# Opportunities and Challenges for ChatGPT and Large Language Models in Biomedicine and Health

Shubo Tian[1], Qiao Jin[1], Lana Yeganova[1], Po-Ting Lai[1], Qingqing Zhu[1], Xiuying Chen[2], Yifan Yang[1], Qingyu Chen[1], Won Kim[1], Donald C. Comeau[1], Rezarta Islamaj[1], Aadit Kapoor[1], Xin Gao[2], Zhiyong Lu[1]*

Affiliations:

[1]National Library of Medicine, National Institutes of Health

[2]King Abdullah University of Science and Technology

*Corresponding Author

## Abstract

ChatGPT has drawn considerable attention from both the general public and domain experts with its remarkable text generation capabilities. This has subsequently led to the emergence of diverse applications in the field of biomedicine and health. In this work, we examine the diverse applications of large language models (LLMs), such as ChatGPT, in biomedicine and health. Specifically we explore the areas of biomedical information retrieval, question answering, medical text summarization, information extraction, and medical education, and investigate whether LLMs possess the transformative power to revolutionize these tasks or whether the distinct complexities of biomedical domain presents unique challenges. Following an extensive literature survey, we find that significant advances have been made in the field of text generation tasks, surpassing the previous state-of-the-art methods. For other applications, the advances have been modest. Overall, LLMs have not yet revolutionized biomedicine, but recent rapid progress indicates that such methods hold great potential to provide valuable means for accelerating discovery and improving health. We also find that the use of LLMs, like ChatGPT, in the fields of biomedicine and health entails various risks and challenges, including fabricated information in its generated responses, as well as legal and privacy concerns associated with sensitive patient data. We believe this survey can provide a comprehensive and timely overview to biomedical researchers and healthcare practitioners on the opportunities and challenges associated with using ChatGPT and other LLMs for transforming biomedicine and health.

# 1. Introduction

The recent release of ChatGPT [1] and the subsequent launch of GPT-4 [2] have captured massive attention among both the general public and domain professionals and has triggered a new wave of development of large language models (LLMs). LLMs such as ChatGPT and GPT-4 are language models that have billions of parameters in model size and are trained with data sets containing tens or hundreds of billions of tokens. They are considered as foundation models [3] that are pretrained on a large scale data and can be adapted to different downstream tasks. LLMs have achieved impressive performance in a wide range of applications in various fields including the biomedical and health domains. A keyword search of "large language models" OR "ChatGPT" in PubMed returned 582 articles by the end of May 2023. Moreover, the number of publications on the topic has grown exponentially from late 2022 and doubled every month in the last six months, covering the technology and its implications for various biomedical and health applications.

Furthermore, several biomedical specific LLMs have been developed by either training from scratch or fine-tuning existing pre-trained LLMs with biomedical data [4-9]. To provide a comprehensive overview to biomedical researchers and healthcare practitioners on the possible and effective utilization of ChatGPT and other LLMs in our domain, we performed a literature survey, exploring their potentials in a wide variety of different applications such as biomedical information retrieval, question answering, medical text summarization, information extraction, and medical education. Additionally, we delve into the limitations and risks associated with these language models.

Specifically, due to the remarkable language generation capabilities, our focus centers on ChatGPT and other LLMs within the domain of generative artificial intelligence (AI). We searched articles containing keywords related to large language models, such as "GPT", "ChatGPT" or "large language model", along with keywords of biomedical applications on PubMed[1], medRxiv[2], arXiv[3] and Google Scholar[4], and included the articles identified as relevant for our review. To the best of our knowledge, this is the first comprehensive survey of opportunities and challenges on ChatGPT and other LLMs for fundamental applications in seeking information and knowledge discovery in biomedicine and health, although there are several previous survey papers on general LLMs [10, 11] and use of ChatGPT for different specific health applications [12-14]. By discussing the capabilities and limitations of ChatGPT and LLMs, we strive to unlock their immense potential in addressing the current challenges within the fields of biomedicine and health. Furthermore, we aim to highlight the role of these models in driving innovation and ultimately improving healthcare outcomes.

---

[1] https://pubmed.ncbi.nlm.nih.gov/
[2] https://www.medrxiv.org/
[3] https://arxiv.org/
[4] https://scholar.google.com/

## 2. Overview ChatGPT and Domain-specific LLMs

### 2.1. Overview of General LLMs

A language model (LM) is a statistical model that computes the (joint) probability of a sequence of words (or tokens). Research on language models has been going on for a long period of time [15]. In 2017 the transformer model introduced by Vaswani et al. [16] became the foundational architecture for most modern language models including ChatGPT. The transformer architecture includes an encoder of bidirectional attention blocks and a decoder of unidirectional attention blocks. Based on the modules used for model development, most recent LMs can be grouped into three categories: encoder-only LMs such as BERT (Bidirectional Encoder Representations from Transformers) [17] and its variants, decoder-only LMs such as the GPT (Generative Pre-trained Transformer) family [18-20] and encoder-decoder LMs such as T5 (Text-to-Text Transfer Transformer) [21] and BART (Bidirectional and AutoRegressive Transformers) [22]. Encoder-only and encoder-decoder language models are usually trained with an infilling ("masked LM" or "span corruption") objective along with an optional downstream task, while decoder-only LMs are trained with autoregressive language models that predict the next token given the previous tokens.

Although the encoder-only and encoder-decoder models have achieved state-of-the-art performance across a variety of natural language processing (NLP) tasks, they have the downside that requires significant amount of task-specific data for fine-tuning the model to adapt to the specific tasks. This process needs to update the model parameters and adds complexity to model development and deployment.

Unlike those models, when GPT-3 [19] was released, it demonstrated that large decoder-only language models trained on large text corpus gained significantly increased capability [23] for natural language generation. After training, the models can be directly applied to various unseen downstream tasks through in-context learning such as zero-shot, one-shot or few-shot prompting [19]. This led to a recent trend toward development of decoder-only LLMs in the following years. Following GPT-3, a number of powerful LLMs such as PaLM [24], Galactica [25], and the most recent GPT-4 [2], have been developed. For more information on these general-domain models, readers are invited to consult [10, 11].

While LLMs are powerful, they are still likely to produce content that is toxic, biased, or harmful for humans since the large corpus used for model training could contain both high-quality and low-quality data. Thus, it is extremely important to align LLMs to generate outputs that are helpful, honest, and harmless for their human users. To achieve this, Ouyang et al. [26] designed an effective approach of fine-tuning with human feedback to fine-tune GPT-3 into the InstructGPT model. They first fine-tuned GPT-3 on a dataset of human-written demonstrations of the desired output to prompts using supervised learning, then further fine-tuned the supervised model through reinforcement learning from human feedback (RLHF). This process was referred as alignment tuning. It was also applied in the development process of ChatGPT and became an effective practice for development of faithful LLMs.

With model size growing bigger, fine-tuning LLMs for downstream tasks becomes inefficient and costly. Alternatively, prompt engineering serves as the key to unlock the power of LLMs given their strong in-context learning ability. As demonstrated by GPT-3, large language models were able to achieve promising performance on a wide range of natural language tasks through in-context learning by prompting that used a natural language instruction with or without demonstration examples as prompt for the model to generate expected outputs. Wei et al. [27] showed that chain-of-thought prompting through a series of intermediate reasoning steps was able to significantly improve LLMs' performance on complex arithmetic, common sense, and symbolic reasoning tasks. As a useful approach, designing prompts suitable for specific tasks through prompt engineering became an effective strategy to elicit the in-context learning ability of LLMs. The process of training, fine-tuning with human feedback, and unlocking power of LLMs through prompt engineering becomes the paradigm of LLMs as shown in Figure 1.

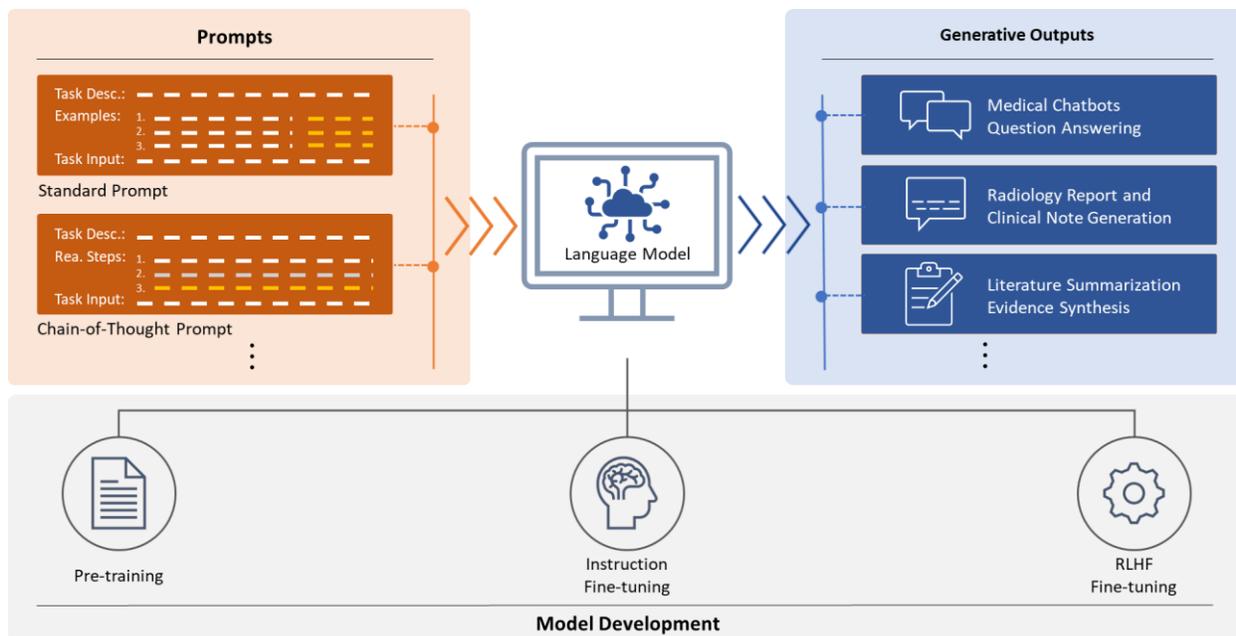

Figure 1. The paradigm of LLMs. Pre-training: LLMs are trained on large scale corpus using autoregressive language model; Instruction Fine-tuning: pre-trained LLMs are fine-tuned on a dataset of human-written demonstrations of the desired output behavior on prompts using supervised learning; RLHF Fine-tuning: a reward model is trained using collected comparison data, then the supervised model is further fine-tuned against the reward model using reinforcement learning algorithm.

## 2.2. LLMs for Biomedical and Health Applications

Development of LLMs has been steadily setting new state-of-the-art performance on a variety of tasks in general NLP as well as in biomedical NLP specifically [8, 26, 28-30]. An example is the performance of LLMs on the MedQA dataset, a widely used biomedical question answering dataset that comprises questions in the style of the US Medical Licensing Exam (USMLE) and is used for evaluation of LLMs' reasoning capabilities. In less than half a year, LLM performance has approached a level close to human expert by Med-PaLM 2 [30] from the level of human passing by GPT-3.5 [31], as depicted in Figure 2. These achievements have been accomplished by adapting the LLMs for biomedical QA through different strategies.

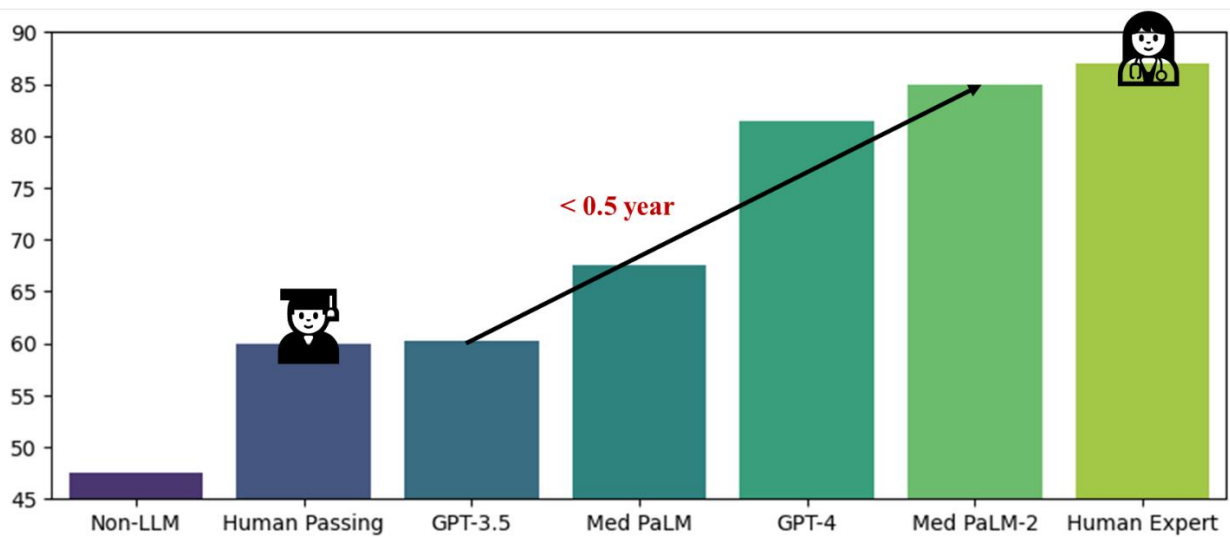

Figure 2. Performance of LLMs vs. human on the MedQA (USMLE) dataset in terms of accuracy.

There are several strategies that can be applied to adapt ChatGPT and other LLMs for specific applications in biomedicine and health. When a large amount of data as well as more computing resources and expertise are available, domain specific language models can be developed by pre-training from scratch or from checkpoints of existing general language models. Alternatively, strategies such as fine-tuning with task specific data, instruction fine-tuning and/or RLHF fine-tuning, soft prompt tuning, and prompt engineering can be employed to adapt existing pre-trained language models to specific domain applications. Explanation of these strategies and corresponding examples are listed as follows.

- **Pre-training from scratch** is to create a specialized language model by pre-training the language model with randomly initialized parameters on a large biomedical corpus using the training objective of either infilling ("masked LM" or "span corruption") or an autoregressive language model. Both BioMedLM [6] and BioGPT [7] are specialized biomedical language models developed by pre-training on a corpus of PubMed articles from scratch.
- **Pre-training from checkpoints of existing general LMs** is to develop a specialized language model by initializing its parameters from the checkpoint of an existing general language model and further pre-training the model on a biomedical corpus with the infilling or autoregressive language model training objectives. The PMC-LLaMA [9] is a model developed by further pre-training the LLaMA-7B [32] model on PubMed Central articles.
- **Fine-tuning with task specific data** has been frequently used to adapt relatively smaller LMs for specific downstream tasks. This strategy is to fine-tune the existing LMs on the training data of a downstream task with the same training objective of the task. The developers of BioGPT [7] also fine-tuned BioGPT on task specific data after it was pre-trained from scratch.

- **Instruction fine-tuning and/or RLHF fine-tuning** is the strategy for aligning LLMs with better instruction responses by fine-tuning the model on data of instruction-response pairs through supervised learning and/or reinforcement learning. Several LLMs including Med-PaLM 2 [30], Clinical Camel [33], ChatDoctor [34], and MedAlpaca [35] have been developed through instruction fine-tuning.
- **Soft prompt tuning** is the learning of soft prompt vectors that can be used as prompts to LLMs for specific downstream tasks. It is a strategy to take advantage of the benefit from gradient-based learning through a handful of training examples while keeping parameters of the LLMs frozen. The model of Med-PaLM [8] is the result of adapting Flan-PaLM [36] to the biomedical domain through soft prompt tuning.
- **Prompt engineering** is the process of designing appropriate prompts to adapt LLMs for specific downstream tasks by leveraging the powerful in-context learning capabilities of LLMs without the need of gradient-based learning. Various prompt engineering techniques have been developed and applied for adapting LLMs to biomedical and health related tasks.

Although previous research has shown that language models pre-trained with biomedical domain specific data can benefit various in-domain downstream tasks [37-39], pre-training a language model from scratch or existing checkpoint can be very costly, especially when the sizes of language models growing larger. Adapting LLMs through instruction fine-tuning, soft prompt tuning and prompt engineering can be more cost effective and accessible. In addition, while the strategies listed above can be employed independently, they can also be applied in combination when applicable.

The advances of LLMs in recent years have led to development of a number of specialized biomedical LLMs such as BioMedLM [6], BioGPT [7], PMC-LLaMA [9], Med-PaLM [8], Med-PaLM 2 [30], Clinical Camel [33], ChatDoctor [34], and MedAlpaca [35]. Uses of general LLMs including GPT-3 [19], GPT-3.5 [31], ChatGPT [1], GPT-4 [29], Flan-PaLM [8], and Galactica [25] for biomedical applications are being extensively evaluated. Table 1 provides a list of domain specific LLMs [6-9, 30, 33-35, 40-46]. Performances of various LLMs on different biomedical application tasks are described in Section 3.

Table 1. Specialized LLMs in biomedical and health fields.

| LLM | Size | Description |
|---|---|---|
| BioMedLM | 2.7B | Developed based on HuggingFace's implementation of the GPT-2 model with small changes through pre-training from scratch on data of 16 million PubMed Abstracts and 5 million PubMed Central full-text articles contained in the Pile dataset [40] |
| BioGPT | 347M and 1.5B | Developed based on the GPT-2 architecture through pre-training from scratch on a corpus of 15 million PubMed articles that have both title and abstract |
| PMC-LLaMA | 7B | Developed by further pre-training from the LLaMA 7B model on 4.9 million PubMed Central articles filtered from the S2ORC Datasets [41] for only 5 epochs |

| Med-PaLM 2 | N/a | Developed by instruction fine-tuning of PaLM 2 [42] based on a data mixture of medical question answering datasets including MedQA [43], MedMCQA [44], HealthSearchQA [8], LiveQA [45] and MedicationQA [46] |
| --- | --- | --- |
| Clinical Camel | 13B | Developed by instruction fine-tuning of the LLaMA 13B model on general multi-step conversations in the ShareGPT[5] data and synthetic dialogues transformed from the MedQA data and clinical review articles |
| ChatDoctor | 7B | Developed by instruction fine-tuning of the LLaMA 7B model on more than 100,000 real-world patient-physician conversations collected from two online medical consultation sites |
| MedAlpaca | 7B and 13B | Developed by instruction fine-tuning of the LLaMA 7B and 13B models on the Medical Meadow data, a collection of reformatted instruction-response pairs including datasets for medical NLP tasks and data crawled from various internet resources |
| Med-PaLM | 540B | Developed by adapting Flan-PaLM to medical domain through instruction prompt tuning on 40 examples |

## 3. Applications of ChatGPT and LLMs in Biomedicine and Health

ChatGPT and other LLMs can be used in a wide range of biomedical and health applications. In this survey, we cover applications that are fundamental in satisfying information needs of clinical decision making and knowledge acquisition, including biomedical information retrieval, question answering, medical text summarization, information extraction, and medical education.

### 3.1. Information Retrieval

Information retrieval (IR) is an integral part in clinical decision making [47] and biomedical knowledge acquisition [48], as it covers various information-seeking behaviors such as literature search [49], question answering [50], and article recommendation [51]. Large language models like ChatGPT hold significant potentials in changing the way people interact with medical information online [52].

First and foremost, current LLMs may not be directly used as a search engine because their output can contain fabricated information, commonly known as the hallucination issue. For example, when prompted "Could you tell me what's the relation between p53 and depression? Please also provide the references by PMIDs.", ChatGPT makes up the content of PMID 25772646 (Perspectives on thyroid hormone action in adult neurogenesis) to support its incorrect answers. This behavior makes retrieving out-of-context knowledge from ChatGPT potentially dangerous by leading the users to draw incorrect conclusions.

---
[5] https://sharegpt.com/

However, LLMs might facilitate the interpretation of traditional IR systems by text summarization. It is also been shown by several pilot studies in biomedicine that when LLMs are provided with enough contexts and background information, they can be very effective at reading comprehension [8, 31] and could generate fluent summaries with high fidelity [53]. These results suggest that ChatGPT might be able to summarize the information returned by a traditional IR system and provide a high-level overview or a direct answer to users' queries. Many search engines have integrated LLMs into their result page. For example, *You.com* and the *New Bing* provide ChatGPT-like interactive agents that are contextualized on the web search results to help users navigate them; *scite.ai* presents LLM-generated summaries with references linked to the retrieved articles for literature search results. While afore-mentioned features are potentially beneficial for all IR systems, researchers have cautioned that the generated outputs must be carefully verified. Although LLMs can summarize in-context information with high fidelity, there is no guarantee that such summaries are error-free [54].

LLMs like ChatGPT can also be used for query enrichment and improving search results with generating more specific queries, expanding a user's search query to include additional relevant terms, concepts, or synonyms that may improve the accuracy and relevance of the search results. For instance, Wang et al. [55] used ChatGPT to formulate and refine Boolean queries for systematic reviews. They created an extensive set of prompts to investigate tasks on over 100 systematic review topics. Their experiments were conducted on two benchmarking collections: the CLEF technological assisted review (TAR) datasets [56-58] and the Systematic Review Collection with Seed studies [59]. The ChatGPT-generated queries were compared to the original queries, the Baseline Conceptual and Objective. Evaluation was performed using precision, recall, and F1 and F3 score metrics. Their results show that the ChatGPT-generated queries have higher precision, but lower recall compared to the queries generated by the current state-of-the-art method [55].

## 3.2. Question Answering

Question answering (QA) denotes the task of automatically answering a given question. In biomedicine, QA systems can be used to assist clinical decision support, create medical chatbots, and facilitate consumer health education [50]. Based on whether the supporting materials are available, QA tasks can be broadly classified into open(-domain) QA and machine reading comprehension. In open QA, only the question is provided (e.g., a consumer health search query), and a model needs to use external or internal knowledge to answer the question. In machine reading comprehension, both the question and the material for answering the question are available, e.g., in the case where doctors ask questions about specific clinical notes.

A wide variety of biomedical QA datasets have been introduced over the past decade, including BioASQ [60, 61], MedMCQA [44], MedQA (USMLE) [43], PubMedQA [62], GeneTuring [63]. MedMCQA and MedQA are general medical knowledge tests in US medical licensing exam (USMLE) and Indian medical entrance exams, respectively. Both datasets are open-domain tasks where only the question and 4-5 answer options are available. In contrast, the questions and answers in GeneTuring are in the genomics domain such as gene name conversion and nucleotide sequence alignment. On the other hand, BioASQ and PubMedQA provide relevant PubMed articles as supporting materials to answer the given question. The biomedical QA tasks are evaluated using the classification accuracy of the possible answers (4-5 provided options for MedMCQA and MedQA (USMLE), entities for GeneTuring, yes/no for BioASQ, and yes/no/maybe for PubMedQA).

Table 2 shows the performance of LLMs on three commonly used biomedical QA tasks. Overall, the best results are achieved by either Med-PaLM 2 (on MedQA and PubMedQA) or GPT-4 (on MedMCQA), which are currently the largest LLMs containing hundreds of billions of parameters. Notably, they achieve comparable performance on the MedQA dataset and higher performance on the PubMedQA dataset in comparison to the human expert. FLAN-PaLM and GPT-3.5 also achieve high scores on PubMedQA, but are much worse than Med-PaLM 2 and GPT-4 on the MedQA and MedMCQA datasets. This is probably because PubMedQA mainly requires the reading comprehension capability (reasoning), while the other open QA datasets require both reasoning and knowledge. However, smaller LLMs (<10B), such as BioMedLM and PMC-LLaMA, perform similarly to DRAGON [64], a BERT-sized SOTA model enhanced by domain knowledge. This suggests that auto-regressive LLMs could scale to large enough model sizes to outperform smaller models augmented by structured domain knowledge.

Table 2. Performance of LLMs on biomedical QA tasks.

| Model | Learning | MedQA (USMLE) | PubMedQA (r.r./r.f.) | MedMCQA (dev/test) |
|---|---|---|---|---|
| Human expert [31] | - | 87.0 | 78.0 / 90.4 | 90.0 |
| Human passing [31] | - | 60.0 | - | 50.0 |
| Med-PaLM 2 [30] | Mixed | 86.5* | 81.8 / - | 72.3* / - |
| GPT-4 [29] | Few-shot | 86.1 | 80.4 / - | 73.7 / - |
| FLAN-PaLM [8] | Few-shot | 67.6 | 79.0 / - | 57.6 / - |
| GPT-3.5 [31] | Few-shot | 60.2 | 78.2 /- | 59.7 / 62.7 |
| Galactica [25] | Mixed | 44.4 | 77.6* / - | 52.9* / - |
| BioMedLM [6] | Fine-tune | 50.3 | 74.4 / - | - |
| BioGPT [7] | Fine-tune | - | - /81.0 | - |
| PMC-LLaMA [9] | Fine-tune | 44.7 | 69.5 / - | - / 50.5 |
| Non-LLM SOTA [64] | Fine-tune | 47.5 | 73.4 / - | - |

Note: All numbers are accuracy in percentages. Underscored numbers denote the highest performance. *Fine-tuning. r.r.: reasoning-required; r.f.: reasoning-free.

Answering biomedical questions requires up-to-date and accurate knowledge. To address the hallucination issue [65] in medical QA systems, one of the current solutions is retrieval augmentation, which refers to the approach of combining LLMs with a search system, such as the New Bing for the general domain and Almanac [66] in the clinical domain. For a given question, the system will first retrieve relevant documents as supporting materials, and then prompt LLMs to answer the question based on the retrieved documents. In this case, LLMs might generate less hallucinations since they are good at summarizing content. However, such systems are still not free from errors [54] and there is a need for more systematic evaluations [52]. Another promising direction for tackling the hallucination issue is to augment LLMs with additional tools [67-70]. For example, the GeneTuring dataset contains information-seeking questions for specific SNPs such as rs745940901. However, auto-regressive LLMs possess no knowledge about that SNP and most commercial search engines return no results to this query, so retrieval-augmentation might not work either. In this case, the information source is only accessible via the NCBI dbSNP database, and augmenting LLMs with NCBI Web database utility APIs can potentially solve the hallucination issue with regard to specific entities in biomedical databases [67].

Consumers have been relying on web search engines like Google for their medical information needs [71]. It is conceivable that they might turn to LLM chatbots because the dialogue interface can directly answer their questions and follow-ups. In fact, there has already been several studies, such as Clinical Camel [33], DoctorGLM [72], ChatDoctor [34], HuaTuo [73], MedAlpaca [35], that attempt to create clinical chatbots by instruction fine-tuning open source LLMs (e.g., LLaMA) on biomedical corpora. However, most of such studies only use small private datasets for evaluation, and the accuracy, generalizability, and actual utility of such dialogue systems remains unclear.

### 3.3. Biomedical Text Summarization

Text summarization in the biomedical and health fields is an important application of natural language processing and machine learning. This process involves condensing lengthy medical texts into shorter, easy-to-understand summaries without losing critical information. Summarization in the medical field can be particularly challenging due to the complexity of the language, terminology, and concepts. In this section, we will introduce three application scenarios for text summarization in biomedicine: literature summarization, radiology report summarization, and clinical note summarization.

The first important application is medical literature summarization [74]. A well-summarized literature review can help in condensing a large volume of information into a concise, readable format, making it easier for readers to grasp the key findings and conclusions. Towards this goal, Cohan et al. [75] introduced a scholar paper summarization task, where they proposed a large-scale dataset of long and structured scientific papers obtained from PubMed, where the abstracts are regarded as the summary of the paper. Pang et al. [76] achieved state-of-the-art performance on this dataset with top-down and bottom-up inference technique. Taking a step from paper summarization to literature summarization, Chen et al. [77] proposed a related work generation task, where the related work section is considered as the literature review for the specific field. With the development of LLMs, it is expected that more related documents can be considered [78], and better evaluation metrics can be proposed to evaluate the quality of summaries [79].

We next examine how summarization techniques can help medical applications such as radiology report summarization [80]. This is the process of condensing lengthy and detailed radiology reports into concise, informative, and easily understandable summaries. Radiology reports contain critical information about a patient's medical imaging results, such as X-rays, CT scans, MRI scans, and ultrasound examinations. Representative datasets include MIMIC-CXR [81], which is a large-scale radiography dataset comprising 473,057 chest X-ray images and 206,563 reports. Hu et al. [80] utilized an anatomy-enhanced multimodal model to achieve state-of-the-art results in terms of the ROUGE and CheXbert [82] metrics. In the era of LLMs, Ma et al. [83] proposed ImpressionGPT, which leverages the in-context learning capability of LLMs for radiology report summarization. Wang and colleagues proposed ChatCAD [84], a framework that summarizes and reorganizes information from radiology report to support query-aware summarization.

Finally, clinical notes summarization [85] aims to summarize other non-radiology clinical notes, which helps doctors and other healthcare professionals quickly grasp the essential information about a patient's condition, treatments, and progress. While radiology report summarizations are more focused, delivering insights based on imaging studies, clinical note summarization involves summarizing the overall status, progress, and plan for a patient based on various clinical observations, examinations, and patient interactions [86]. McInerney et al. proposed and evaluated models that extract relevant text snippets from patient records to provide a rough case summary [87]. Recently, Peng et al. demonstrated that while ChatGPT can condense pre-existing systematic reviews, it frequently overlooks crucial elements in the summary, particularly failing to mention short-term or long-term outcomes that are often associated with different levels of risk [88]. Patel and Lam [89] discussed the possibility of using an LLM to generate discharge summaries, and Tang et al. [90] tested performance of ChatGPT on their in-house medical evidence dataset. As a concluding work, Ramprasad et al. discussed the current challenges in summarizing evidence from clinical notes [91].

## 3.4. Information Extraction

Information extraction involves extracting specific information from unstructured biomedical text data and organizing the extracted information into structured format. The two most studied IE tasks are (a) named entity recognition (NER): recognizing biological and clinical entities (e.g., diseases) asserted in the free text; and (b) relation extraction (RE): extracting relations between entities in the free text.

Pre-trained language models have been widely used in NER and RE methods. The encoder-only LMs such as BERT are typically fine-tuned with annotated data via supervised learning before being applied for NER and RE tasks. Instead, using decoder-only LMs for NER and RE will usually model them as text generation tasks to directly generate the entities and the relation pairs. Current state-of-the-art (SOTA) NER and RE performance were mostly achieved by models based on encoder-only LMs that were pre-trained on biomedical and clinical text corpus [92, 93] or machine learning method [94].

Recently, several studies have been conducted to explore the use of GPT-3 and ChatGPT for biomedical NER and RE tasks. For example, Agrawal et al. [95] used GPT-3 for NER task on the CASI dataset and showed that GPT-3 was able to outperform the baseline model by observing a single input-output pair. Caufield et al. [96] developed SPIRES by recursively querying GPT-3 to obtain responses and achieved an F1-score of 40.65% for RE on the BC5CDR dataset [94] using zero-shot learning without fine-tuning on the training data. Gutiérrez et al. [97] used 100 training examples to explore GPT-3's in-context learning for biomedical information extraction and discovered that GPT-3 outperformed PubMedBERT, BioBERT-large, and ROBERTa-large in few-shot settings on several biomedical NER and RE datasets. A benchmark study conducted by Chen et al. [98] employed prompt engineering method to evaluate ChatGPT's performance on biomedical NER and RE in the BLURB benchmark datasets including BC5CDR-chemical [94], BC5CDR-disease [94], NCBI-disease [99], BC2GM [100], JNLPBA [101], ChemProt [102], DDI [103], and GAD [104] in a zero-shot or few-shot manner. Chen et al. [105] performed a pilot study to establish the baselines of using GPT-3.5 and GPT-4 for biomedical NER and RE at zero-shot and one-shot settings. They selected 180 examples with entities or relations and 20 examples without entities or relations from each of the BC5CDR-chemical, NCBI-disease, ChemProt and DDI datasets and designed consistent prompts to evaluate the performance of GPT-3.5 and GPT-4. Table 3 and Table 4 summarize performance of different LMs on some commonly used NER and RE benchmark datasets.

Table 3. Performance of LLMs for NER compared to SOTA on selected datasets (F1-score in %).

| Language Model | Method | BC2GM | BC5CDR-chemical | BC5CDR-disease | JNLPBA | NCBI-disease |
|---|---|---|---|---|---|---|
| SOTA | Task fine-tuning | 84.52 | 93.33 | 85.62 | 79.10 | 87.82 |
| GPT-3 | Few-shot | 41.40 | 73.00 | 43.60 | | 51.40 |
| GPT-3.5 | Zero-shot | | 29.25 | | | 24.05 |
| | One-shot | | 18.03 | | | 12.73 |
| ChatGPT | Zero-shot or few-shot | 37.54 | 60.30 | 51.77 | 41.25 | 50.49 |
| GPT-4 | Zero-shot | | 74.43 | | | 56.73 |
| | One-shot | | 82.07 | | | 48.37 |

Table 4. Performance of LLMs for RE compared to SOTA on selected datasets (F1-score in %).

| LM | Method | BC5CDR | CHEMPROT | DDI | GAD |
|---|---|---|---|---|---|
| SOTA | Task fine-tuning | 57.03 | 77.24 | 82.36 | 83.96 |
| BioGPT | Task fine-tuning and few-shot | 46.17 | | 40.76 | |
| GPT-3 | Few-shot | | 25.90 | 16.10 | 66.00 |
| SPIRES | Zero-shot | 40.65 | | | |
| GPT-3.5 | Zero-shot | | 57.43 | 33.49 | |
| | One-shot | | 61.91 | 34.40 | |
| ChatGPT | Zero-shot or few-shot | | 34.16 | 51.62 | 52.43 |
| GPT-4 | Zero-shot | | 66.18 | 63.25 | |
| | One-shot | | 65.43 | 65.58 | |

The drawback of models achieving SOTA NER and RE performance is their need of labeled data. The remarkable in-context learning abilities of LLMs such as ChatGPT exhibited great potential and provided significant advantages for biomedical NER and RE in circumstances where labeled data are not available. However, they are still not able to surpass performance of LMs that are fine-tuned on task specific datasets. In addition, several challenges still exist in the use of ChatGPT and other LLMs for information extraction. The generative outputs of ChatGPT and other LLMs sometimes will re-phrase the identified entities or predicted relations which make them difficult to verify. ChatGPT and other LLMs can also produce entities and relations that sound plausible but not factually true. Searching for prompts that are appropriate for NER and RE can be challenging as well. Given all these challenges, extensive research is needed to explore effective approaches to leverage ChatGPT and other LLMs for biomedical information extraction.

## 3.5. Medical Education

The use of LLMs in medical education is an exciting and rapidly growing area of research and development. In particular, LLMs have a potential to mature into education applications and provide alternative learning avenues for students to help them acquire and retain knowledge more efficiently.

One of the attractive features of ChatGPT is its ability to interact in a conversational way [106]. The dialogue format makes it possible for ChatGPT to answer follow-up questions and communicate in a conversational format. An early application of ChatGPT in education is a pilot study conducted by Khan academy [107]. Although the application is not in healthcare education, but in general education for students in grades K-12, it is an illustration of the model's integration in educational environment. Khanmigo, a real-time chat bot, analyzes the answers, and guides the student towards the solution by asking questions and providing encouragement.

In addition, ChatGPT is equipped with the ability to provide insights and explanations, suggesting that LLMs may have the potential to become interactive medical education tools to support learning. One of the features making them suitable for education is their ability to answer questions and provide learning experiences for individual students, helping them learn more efficiently and effectively.

ChatGPT can also be used for generating case scenarios [108, 109] or quizzes to help medical students practice and improve their diagnostic and treatment planning abilities [110]. For example, the author in [109], engages in a dialog with a chatbot, asking it to simulate a patient with undiagnosed diabetes and the common labs that may need to be run.

LLMs can also be used to help medical students improve their communication skills. By analyzing natural language inputs and generating human-like responses, LLMs can help students practice their communication skills in a safe and controlled environment. For example, an LLM might be used to simulate patient interactions, allowing students to practice delivering difficult news or explaining complex medical concepts in a clear and concise manner.

## 4. Limitations and Risks of LLMs

While LLMs like ChatGPT demonstrate powerful capabilities, these models are not without limitations. In fact, the deployment of LLMs in high-stakes applications, particularly within the biomedical and health domain, presents challenges and potential risks. The limitations, challenges, and risks associated with LLMs have been extensively discussed in previous research [19, 111, 112], and in this survey, we will specifically focus on those relevant to the context of biomedical and health domains.

## 4.1. Hallucination

All language models have the tendency to hallucinate – produce content that may seem plausible but is not correct. When such content is used for providing medical advice or in clinical decision making the consequence can be particularly harmful and even disastrous. The potential danger associated with hallucinations can become more serious as the capabilities of LLMs continue to advance, resulting in more convincing, persuasive and believable hallucinations. These systems are known to lack transparency – inability to relate to the source, which creates a barrier for using the provided information. For health care professionals to use LLMs in support of their decision making, great caution should be exercised to verify the generated information.

Another concern is that LLMs may not be able to capture the full complexity of medical knowledge and clinical decision-making or produce erroneous results. While LLMs can analyze vast amounts of data and identify patterns, they may not be able to replicate the nuanced judgment and experience of a human clinician. The usage of non-standard terminologies presents an additional complication.

## 4.2. Fairness and Bias

In recent years, fairness has gained the attention of ML research communities as a crucial consideration for both stable performance and unbiased downstream prediction. Many studies have shown that language models can amplify and perpetuate biases [2, 113] because they learned from historical data. This may inadvertently perpetuate biases and inequalities in healthcare. In a recent study, researchers show that text generated with GPT-3 can capture social bias [114]. Although there exists a lot of research in the general domain regarding fairness in ML and NLP including gender and racial bias, little work has been done in biomedical domain. Many current datasets do not have demographic information, as this relates to privacy concerns in medical practices. Unfair and biased model in the biomedical and health domain can lead to detrimental outcomes and affect the quality of treatment a patient receives [115-117].

## 4.3. Privacy

The corpora used for LLMs training usually contain a variety of data from various sources, which may include private personal information. Huang et al. [118] found that LMs can leak personal information. It was also reported that GPT-4 has the potential to be used for attempt to identify private individuals and associate personal information such as geographic location and phone number [112]. Biomedical and clinical text data used for training LLMs may contain patient information and pose serious risks to patient privacy. LLMs deployed for biomedical and health applications can also present risks to patient privacy, as they may have access to patient characteristics, such as clinical measurements, molecular signatures, and sensory tracking data.

## 4.4. Legal and Ethical Concerns

Debates on legal and ethical concerns of using AI for medicine and healthcare have been carried out continuously in recent years [119]. The widespread interests in ChatGPT recently also raised many concerns on legal and ethical issues regarding the use of LLMs like ChatGPT in medical research and practices [120, 121]. It was advocated to establish a robust legal framework encompassing transparency, equity, privacy, and accountability. Such framework can ensure safe development, validation, deployment, and continuous monitoring of LLMs, while taking into account limitations and risks [122].

The acknowledgement of ChatGPT as an author in biomedical research has been particularly identified as an ethical concern. Biomedical researchers may have their opinions on whether ChatGPT or other LLMs should be welcomed to their ranks. In fact, several papers already list ChatGPT as an author [123-126]. However, after several ethical concerns were raised [127], several of these papers ended up removing ChatGPT from the author list [128, 129]. One of the problems of allowing machine written articles is whether they could be reliably recognized. In one report, humans detected 68% of generated abstracts, but also flagged 14% of human abstracts as machine generated [130].

The most valid criticism of LLM-assisted generation of scientific papers is accountability. There are no consequences to the LLM if the output is wrong, misleading, or otherwise harmful. Thus, they cannot take responsibility for writing the article [131-135]. Another issue is copyright – in many jurisdictions, machine-generated material may not receive a copyright [120, 131, 136], which poses an obvious problem for journals.

Questions also arise regarding the disclosure of LLM usage during a project or in preparing a paper [133, 135]. There is a long-standing tradition of reporting tools that were used for a project. On one side, if the LLM had a material impact on the study, it should be reported. On the other side, we do not report the spell checker that was used to prepare a paper. Should we report the LLM? Distinguishing these extremes in the context of an LLM will require time and experience.

## 4.5. Lack of Comprehensive Evaluations

LLMs must be comprehensively evaluated with regard to their performance, safety, and potential bias before any implementations in biomedicine. However, evaluating these biomedical LLMs is not trivial. While some traditional NLP tasks such as NER and RE have reliable automatic evaluation metrics like F1 scores, users mostly use LLMs to get free-text response for their biomedical information needs, such as question answering and text summarization. Generally, expert evaluations of such free-text LLM outputs are considered as the gold standard, but getting such evaluations is labor-intensive and not scalable. For example, a panel of clinicians were employed to evaluate the Med-PaLM answers to medical questions among several axes such as scientific consensus, content appropriateness, extent of possible harm, etc. [8]. However, only 140 questions have been evaluated in the study, probably due to the high cost of expert annotations. Another issue for manual evaluation is that there is no consensus on what axes should be evaluated or the scoring guidelines, so the manual evaluation results from different studies are not directly comparable. Therefore, it is imperative to arrive at a reporting consensus, such as the PRISMA statement [137] for systematic reviews, for evaluating biomedical LLMs.

Alternatively, there are two main approaches to evaluate LLM answers without involving expert evaluators. The most common practice is to convert the task into USMLE-style multi-choice questions (such as MedQA, PubMedQA, and MedMCQA) and evaluate the accuracy of LLM-generated answer choices. These tasks serve as a good proxy for evaluating the knowledge reasoning capabilities of LLMs. However, they are not realistic since the answer choices will not be provided in real-life user questions in biomedicine. The other solution is to evaluate the LLM-generated response against a reference answer or summary with automatic metrics. These automatic scoring can be based on lexical overlap such as BLEU [138], ROUGE [139], and METEOR [140], as well as semantic similarity like BERTScore [141], BARTScore [142], and GPTScore [143]. Although these automatic metrics can evaluate free-text LLM outputs in a large scale, they often do not strongly correlate with human judgements [79, 90]. As such, it is also vital to design new evaluation metrics, potentially with LLMs, that can be both scalable and accurate.

## 5. Discussion and Conclusion

In this survey, we reviewed the recent progress of LLMs with a focus on generative models like ChatGPT and their applications in the biomedical and health domains. We find that biomedical and health applications of ChatGPT and other LLMs are being extensively explored in the literature and that some domain specialized LLMs have been developed. Performance of specialized and non-specialized LLMs for biomedical applications have been evaluated on a variety of tasks. Our findings also revealed performance of LLMs varies on different biomedical downstream tasks. LLMs were able to achieve new state-of-the-art performance on text generation tasks such as medical question answering. However, they still underperformed the existing fine-tuning approach of smaller LMs for information extraction.

Looking forward, the opportunities for LLMs present promising prospects for deployment of LLM powered systems for biomedical and health applications in real-life scenarios. In the era of large language models, the future direction of medical summarization holds significant promise. We can anticipate that LLMs will be increasingly utilized to automatically summarize extensive medical literature, radiology reports, and clinical notes. This would facilitate quicker access to vital information and support decision-making processes for healthcare professionals. Additionally, they are expected to better handle complex medical terminology and context, thus improving the quality of summaries. Another potential area of growth is in patient communication. LLMs could be used to transform complex medical jargon into layman's terms, aiding patients in understanding their health conditions and treatment options. Furthermore, a medical classroom furnished with LLMs can bring students more personalized learning experiences and more focus on study of critical thinking and problem-solving skills. A clinical system integrated with LLMs can beneficially provide patients and physicians with efficient and quality health care services through accurate diagnosis, precision medicine, appropriate decision-making, and proper clinical documentation in preparing succinct clinical reports, concise clinical notes, and warm-hearted patient letters.

As a matter of fact, a number of articles have been published on perspectives of using ChatGPT for biomedical and clinical applications in practice [120, 121]. And many experiments have been conducted to evaluate use of ChatGPT in various scenarios in biomedical and clinical workflows. However, until now no actual deployment of ChatGPT or any other LLMs has been reported. Because of the high-stakes nature of biomedical and health settings, deployment of LLMs like ChatGPT into practice requires more prudence given their limitations and risks. In particular, the transparency challenge that the training data of ChatGPT and other LLMs remain closed source increases difficulties in the evaluation of LLMs.

While the potential benefits are immense, we must also be mindful of the risks and challenges as discussed previously. Strategies and techniques need to be developed and deployed for overcoming the limitations of LLMs. To alleviate generation of nonsensical or harmful content, retrieval augmentation techniques can be used, effective prompts need to be crafted, and rigorous evaluation methods shall be applied [67, 144]. To mitigate bias and improve fairness, training data need to be diversified, bias and fairness of LLMs shall be analyzed, and bias detection shall be implemented. To protect privacy of individuals, sensitive personal information shall be limited and deidentified when they are used in LLMs. Regulations shall be created and issued to secure legal and proper use of LLMs. The research community is working hard on development of such strategies and techniques. Ensuring the ethical use of AI in healthcare, maintaining patient privacy, mitigating biases in AI models, and increasing transparency of AI models are some of the significant considerations for future development in this area. Therefore, a multidisciplinary approach, including input from healthcare professionals, data scientists, ethicists, and policymakers, will be crucial to guide the future direction of future research and development in the era of LLMs.

## Acknowledgement

This research was supported by the NIH Intramural Research Program (IRP), National Library of Medicine. Qingyu Chen was also supported by the National Library of Medicine of the National Institutes of Health under award number 1K99LM014024.